\documentclass[aps,pra,fleqn,twoside,twocolumn,showpacs]{revtex4} 

\usepackage[cp1251]{inputenc}
\usepackage[english]{babel}
\usepackage{amsmath}
\usepackage[]{graphicx}

\begin{document}
\title[Stimulated Raman adiabatic passage]
{STIMULATED RAMAN ADIABATIC PASSAGE\\ IN THE FIELD OF FINITE DURATION PULSES}%
\author{M.V.~GROMOVYI, V.I.~ROMANENKO, L.P.~YATSENKO}
\affiliation{Institute of Physics, Nat. Acad. of Sci. of Ukraine,\\46, Nauky Ave., Kyiv 03680, Ukraine}
\email{vr@iop.kiev.ua}
\thanks{\mbox{}\newline{}\raggedright Ukrainian Journal of Physics, Vol. 54 no.11, pp. 1077-1088 (\texttt{www.ujp.bitp.kiev.ua})}




\begin{abstract}
The theory of stimulated Raman adiabatic passage in a three-level
$\Lambda $-scheme of the interaction of an atom or molecule with light,
which takes the nonadiabatic processes at the beginning
and the end of light pulses into account, is developed.
\end{abstract}


\maketitle

\section{Introduction}

Adiabatic processes in atomic physics, owing to their stability with respect
to a variation of parameters that describe the interaction with a field, play
an especial role as a tool for manipulating atomic and molecular states. For
instance, a fast adiabatic passage of the light pulse carrier frequency
through the resonance with the atomic transition frequency allows one to obtain an atom or
molecule in the excited state with a probability close to
1 \cite{Sho90,Sho08}. In this work, we study the stimulated Raman
adiabatic passage (STIRAP), which can be realized in a three- and a multilevel
scheme of the interaction between the atom and the field and was predicted as
early as in the 1980s \cite{Ore84,Gau88}. The physical
aspects and numerous applications of STIRAP in various branches of physics and
chemistry are discussed in reviews \cite{Ber98,Vit01} and in work
\cite{Sho08}. Recently, STIRAP has been demonstrated to occur in solids
\cite{Got06}.

\begin{figure}[b]
\includegraphics[width=7.0cm]{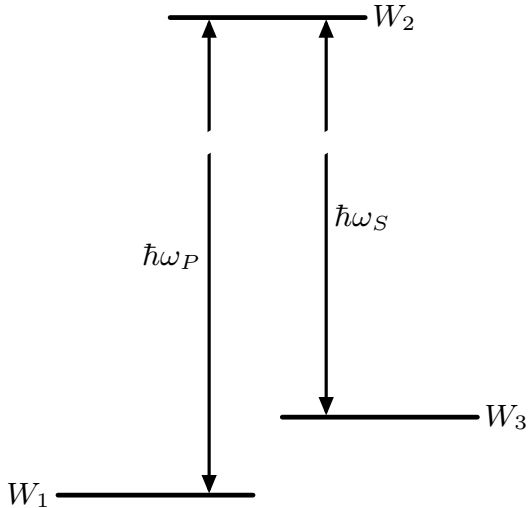}
\vskip-3mm\caption{Scheme of the atom-field interaction. A Stokes
pulse with the carrier frequency $\omega_{S}$ affects the atom
firstly and is followed by the second pumping pulse with the carrier
frequency $\omega_{P}$, by partially overlapping in time with the
first pulse. Here, $W_{1}$, $W_{2}$, and $W_{3}$ are the energies of
atomic states $\left\vert 1\right\rangle $, $\left\vert
2\right\rangle $, and $\left\vert 3\right\rangle $, respectively }
\end{figure}

STIRAP is based on the existence of a trapped or \textquotedblleft
dark\textquotedblright\ state which arises provided that there is a
two-photon resonance between an atom (in what follows, when speaking
about an atom, we also mean a molecule) and a radiation field
produced by two lasers \cite{Alz76,Ari76,Gra78}. In the simplest
case considered in this work, the laser carrier frequencies are so
selected that they are close to the frequency of the transition
between the excited state and two states with lower energies--both
metastable or stable and metastable ones (the three-level
$\Lambda$-scheme of the interaction between the atom and laser
radiation). The radiation of one of the lasers -- the pumping field
-- couples states $\left\vert 1\right\rangle $ and $\left\vert
2\right\rangle $ (see Fig.~1). The field of the other laser -- the
Stokes field -- couples states $\left\vert 3\right\rangle $ and
$\left\vert 2\right\rangle $. The difference between the frequencies
of those fields coincides with the transition frequency $\left\vert
1\right\rangle \leftrightarrow\left\vert 3\right\rangle $. In this
case, some time after the interaction of the atom with the field has
started, the probability to find the atom in the excited state
$\left\vert 2\right\rangle $ is close to zero, i.e. the atomic state
is described by a linear superposition of the basic, $\left\vert
1\right\rangle $, and metastable, $\left\vert 3\right\rangle $,
states. In this case, the populations of the states $\left\vert
1\right\rangle $ and $\left\vert 3\right\rangle $ are determined by
the ratio between the intensities of Stokes and pumping fields.
Being subjected to the action of the pumping field only, the atom is
in state $\left\vert 3\right\rangle $, whereas if only the Stokes
field acts upon the atom, it is in state $\left\vert 1\right\rangle
$. If the intensity ratio changes slowly, the atom transits either
from state $\left\vert 1\right\rangle $ into state $\left\vert
3\right\rangle $ or \textit{vice versa}.

We adopt that, before the interaction between the atom and the field
has started, the former is in state $\left\vert 1\right\rangle $. In
this case, in order to transfer the population from state
$\left\vert 1\right\rangle $ to state $\left\vert 3\right\rangle ,$
it is necessary that a \textquotedblleft
counterintuitive\textquotedblright\ sequence of pulses should affect
the atom. Namely, the atom is first affected by the Stokes pulse and
then by the pumping one which partially overlaps in time with the
Stokes pulse. It is essential that, during the whole time interval
with the atom-field interaction, the population of state $\left\vert
2\right\rangle $ is very small, and the population losses owing to
the spontaneous radiation emission from this state are
insignificant. As a result, STIRAP provides the population transfer
between selected levels with a probability close to 1. In addition,
owing to the adiabaticity of the process (a slow variation of
atom-field interaction parameters), the probability of population
transfer practically does not depend on wide-range variations of the
shape and the intensity of light pulses.

It is natural that the study of the influence of a nonadiabaticity, which
reduces the population transfer probability, on STIRAP has always drawn
attention of researchers. First of all, it should be noted that the
adiabaticity criterion for the process of interaction between the atom and the
field was analyzed in practically every work, where STIRAP was considered
(see, e.g., works \cite{Ore84,Gau88,Ber98,Vit01}). The criterion is based on
the requirement that the eigenvectors of a Hamiltonian that describes
the atom-field interaction should vary slowly in comparison with the difference
between its eigenvalues \cite{Mes79}. Numerically, the adiabaticity is
characterized by the parameter $\varepsilon$ which is reciprocal to the light
pulse area. With the reduction of the adiabaticity parameter $\varepsilon$,
the population $n_{3}(t)$ of the target state $\left\vert 3\right\rangle $ grows,
in general, at $t\rightarrow\infty$ in the course of STIRAP (small
oscillations are possible with a frequency of the order of the Rabi frequency of light pulses).

In the most complete way, the dependence of $n_{3}(\infty)$ on
$\varepsilon$ can be monitored in some cases where the shape of a
pulse envelope allows an exact expression for the population of
atomic states to be found \cite{Car90,Lai96}. Provided that there
are no losses in the atomic state population through the spontaneous
radiation emission -- this condition was postulated in the works
cited above -- the difference of the population in the target state
$\left\vert 3\right\rangle $ from 1 tends to zero, with a reduction
of $\varepsilon$, following different laws for different shapes of
light pulses. For instance, if the amplitudes of the Stokes pulse at
$t\rightarrow-\infty$ and the pumping one at $t\rightarrow\infty$ do
not vanish, and those pulses can be described by analytical
functions within the time interval $[-\infty,\infty]$
\cite{Lai96,Elk95}, then $1-n_{3}\sim \exp\left(
-\mu/\varepsilon\right)$, where $\mu$ is a certain constant of the
order of 1. The dependence of such a type for the given class of
functions has a general character, and the theory developed by
Dykhne \cite{Dyk61} and Davis and Pechukas \cite{Dav76}, which
describes the law, following which the system tends to the adiabatic
state with the reduction of $\varepsilon$, is applicable to those
functions. Another type of dependence -- a power-law with
$1-n_{3}\sim\varepsilon^{2}$ -- was found for pulses with a special
shape, which allows an exact solution of the Schr\"{o}dinger
equation with nonzero first derivatives of the pumping pulse field
at its
beginning and of the Stokes pulse field at the moment of its termination to be obtained \cite{Car90,Lai96}%
. This result is based on a discontinuity of the derivative of the field at the
time moment of its switching-on \cite{Lai96}, that is characteristic of other
quantum-mechanical systems as well, in which the almost adiabatic evolution of the wave function
is possible \cite{Gar62,San66,Yat04}. In the general
case of the nonzero $n$-th derivative at the beginning of a light pulse, taking the
results of the cited works into account, one may expect that $1-n_{3}%
\sim\varepsilon^{2n}$ at $\varepsilon\rightarrow0$ in the STIRAP
case.\looseness=1

While considering finite-duration pulses, we proceed from the
fact that it is the only class of pulses which can be realized under real
experimental conditions. We will examine the cases where the pulse damping
can be neglected (short light pulses) and when the light pulse duration $\tau$
considerably exceeds the inverse lifetime of an atom in the excited state,
$\gamma^{-1}$. The former case was already analyzed for light pulses with
identical amplitudes and with a special pulse shape that allowed the
Schr\"{o}dinger equation to be solved analytically \cite{Car90,Lai96}. The
latter case was analyzed earlier without making allowance for a nonadiabaticity
associated with a jump of the field derivative at the beginning of the Stokes
pulse \cite{Fle96,Rom97}.

For short light pulses ($\gamma\tau\ll1$) and the field-atom interaction close to
the adiabatic one ($\varepsilon\ll1$), we will find the target state population to an
accuracy of $\varepsilon^{2}$ for pulses with arbitrary shape, whose intensity
grows in time at their beginning proportionally to $t^{2}$, and to an
accuracy of $\varepsilon^{4}$, if the intensity at the pulse beginning grows
as $t^{4}$. For long light pulses ($\gamma\tau\gg1$), we will demonstrate
that the transient processes, which arise at the beginning of Stokes pulse action
on the atom, though do not change considerably the probability of population transfer
during STIRAP in comparison with the results obtained in
works \cite{Fle96,Rom97}, do insert substantial corrections into them.

\section{Basic Equations}

Consider an atom, the interaction of which with the field of two light pulses
is described by the three-level scheme (Fig.~1). The pumping pulse denoted by
the subscript $P$ below partially overlaps in time with the Stokes pulse (the
subscript $S$). The Stokes pulse acts upon the atom firstly. The carrier
frequency of the pumping pulse $\omega_{P}$ is identical to that of the transition
between states $\left\vert 1\right\rangle $ and $\left\vert 2\right\rangle $,
and the carrier frequency of the Stokes pulse $\omega_{S}$ to that of the
transition between states $\left\vert 3\right\rangle $ and $\left\vert
2\right\rangle $:
\[
{\mathbf{E}}={\textstyle\frac{1}{2}}{\mathbf{E}}_{{P}}(t){}e^{-i\omega
_{P}t-i\varphi_{{P}}(t)}+{\textstyle\frac{1}{2}}{\mathbf{E}}_{{S}}%
(t){}e^{-i\omega_{S}t-i\varphi_{{S}}(t)}+\mbox{c.c.}
\]
We suppose that the amplitudes of the Stokes, ${\mathbf{E}}_{{P}}(t)$, and
pumping, ${\mathbf{E}}_{{S}}(t)$, pulse fields change smoothly in time with a
characteristic scale comparable to the pulse duration $\tau$.

State $\left\vert 1\right\rangle $, in which the atom stays before its
interaction with the field, is considered to be stable or metastable, whereas
state $\left\vert 3\right\rangle $, into which we intend to transfer the atom
using the STIRAP process, metastable, so that the variations of populations in
those states within the time intervals comparable with the pulse duration,
which take place owing to the processes of spontaneous emission from
them, are neglected. Concerning the spontaneous emission from the excited
state $\left\vert 2\right\rangle $, we suppose that, in the course of this
process, the atom transits into other states different from $\left\vert
1\right\rangle $ and $\left\vert 3\right\rangle $ at the rate $\gamma$
(the lifetime of an atom in the excited state is $\tau_{sp}=\gamma^{-1}$). In
this case, the atomic state can be described by a wave function, the time
evolution of which is described by the Schr\"{o}dinger equation
\begin{equation}
i\hbar\frac{\partial\Psi(t)}{\partial t}=\mathrm{H}(t)\Psi(t) \label{Sch}%
\end{equation}
with a non-Hermitian Hamiltonian that contains $\gamma$ \cite{Sho90}. For
resonant interaction between the atom and the field,
\[
\omega_{P}=(W_{2}-W_{1})/\hbar,\qquad\omega_{S}=(W_{2}-W_{3})/\hbar.
\]
The matrix representation of the atom-field interaction Hamiltonian in the
rotating-wave approximation \cite{Sho90,Sho08} and in the dipole approximation for
the electric field strength looks like \
\begin{equation}
\mathrm{H}(t)=\frac{\hbar}{2}\left[
\begin{array}
[c]{ccc}%
0 & \Omega_{P}(t) & 0\\
\Omega_{P}(t) & -i\gamma & \Omega_{S}(t)\\
0 & \Omega_{S}(t) & 0\\
&  &
\end{array}
\right]  , \label{Ham}%
\end{equation}
and the state vector is a column of probability amplitudes $c_{n}(t)$ to find the atom in
the state $\psi_{n}(t)=\exp(-iW_{n}t/\hbar)\left\vert n\right\rangle $
($n=1,2,3$):
\begin{equation}
\Psi=\left[  c_{1}(t),c_{2}(t),c_{3}(t)\right]  ^{T}. \label{Psi}%
\end{equation}
Here, $\Omega_{P}(t)=-{\mathbf{d}}_{12}{\mathbf{E}}_{P}(t)/\hbar$ and
$\Omega_{S}(t)=-{\mathbf{d}}_{32}{\mathbf{E}}_{S}(t)/\hbar$ are the Rabi
frequencies of the pumping and Stokes pulses, respectively; and $\mathbf{d}$
is the operator of atomic dipole moment. Without any loss of generality, the
Rabi frequencies are considered to be real-valued \cite{Sho90}. We also assume
that, at the moment $t_{i}$, when the pumping pulse starts to affect the atom,
the latter is in state $\left\vert 1\right\rangle $, i.e. the initial
conditions look like
\begin{equation}
c_{1}(t_{i})=1,\qquad{}c_{2}(t_{i})=0,\qquad{}c_{3}(t_{i})=0. \label{ini}%
\end{equation}

It is convenient to characterize the variation of the ratio between the Rabi
frequencies of the Stokes and pumping pulses in time by the time dependence of
the mixing angle \cite{Ber98,Vit01}
\begin{equation}
\vartheta(t)=\mathop{\mathrm{arctan}}\frac{\Omega_{P}(t)}{\Omega_{S}%
(t)}.\label{mix-angle}%
\end{equation}
Then, Hamiltonian (\ref{Ham}) looks like
\begin{equation}
\mathrm{H}(t)=\frac{\hbar}{2}\left[  \arraycolsep=2pt%
\begin{array}
[c]{ccc}%
0 & \Omega(t)\cos\vartheta & 0\\
\Omega(t)\cos\vartheta(t) & -i\gamma & \Omega(t)\sin\vartheta(t)\\
0 & \Omega(t)\sin\vartheta(t) & 0\\
&  &
\end{array}
\right]  \!,\label{Ham-theta}%
\end{equation}
where
\begin{equation}
\Omega(t)=\sqrt{\Omega_{P}(t)^{2}+\Omega_{S}(t)^{2}}.\label{Rabi}%
\end{equation}
The adiabaticity parameter $\varepsilon$, the reduction of which corresponds
to the approach of the atom-field interaction to the adiabatic one, can be
estimated as $\varepsilon\sim(\max[\Omega(t)]\tau)^{-1}$.

\section{Time Dependences of Light Pulse Envelopes}

We illustrate the accuracy of the results obtained below, which describe
the dependence of the population of state $\left\vert 3\right\rangle $ on the
parameters of light pulses by comparing them with the results of numerical
calculations for pulses with model shapes. Let the time dependence of the pumping
pulse repeat that of the Stokes one, but with the delay $t_{d}$:
\begin{equation}%
\begin{split}
&  \Omega_{P}(t)=\Omega_{P0}{F}_{n}(t-t_{d}/2),\\
&  \Omega_{S}(t)=\Omega_{S0}{F}_{n}(t+t_{d}/2),
\end{split}
\label{envelope}%
\end{equation}
where $n=1,2,3\ldots$ enumerates the sequence of the functions
\begin{equation}
{F}_{n}(t)=%
\begin{cases}
\cos^{n}(\pi{}t/\tau), & \text{if $|t|<\tau/2$;}\\
0, & \text{if $|t|\geq\tau/2$}.
\end{cases}
\label{cos}%
\end{equation}
For such light pulses, the atom interacts with the field created by both
pulses from the time moment $t_{i}=-(\tau-t_{d})/2$ till the time moment
$t_{f}=(\tau-t_{d})/2$, i.e. during the time interval $\tau-t_{d}$. If
$t_{d}>\tau$, the Stokes and pumping pulses do not overlap in time.

Among the whole family of light pulses (\ref{envelope}), we use two
envelopes, with $n=1$ and $n=2$. In the first case, the electric field
strength has jumps of its first derivative at the beginning and the end of
pulses; in the latter case, these are jumps of the second derivative. In addition,
the first case for pulses with identical amplitudes, $\Omega_{P0}=\Omega
_{S0}=\Omega_{0}$, and a delay $t_{d}=\tau/2$ between them is remarkable in
that the Rabi frequency $\Omega(t)=\Omega_{0}$ does not depend on time during
the time interval $-\tau/4\leq t\leq\tau/4$, when both pulses interact with
the atom simultaneously, and the mixing angle is a linear function of time,
$\vartheta(t)=\pi t/\tau+\pi/4$ \cite{Lai96}. This feature of the model makes
it possible to obtain analytical expressions for integrals which appear in
the theory and to use a simple example to illustrate the results obtained. The
second model---it was applied earlier, e.g., in work \cite{Yat02}---is close
to Gaussian-like pulses which are often used for the simulation of a light
pulse shape in theoretical calculations \cite{Got06,Lai96,Vit01,Vit97}.

\section{Stimulated Raman Adiabatic Passage in the Field of Short Light
Pulses}

In the case of short light pulses, the duration of which satisfies the
condition $\gamma\tau\ll1$, the term in Hamiltonian (\ref{Ham-theta}) which
describes the relaxation can be neglected. Let us pass to the basis of
characteristic (adiabatic) states of Hamiltonian (\ref{Ham-theta}). They
satisfy the equation
\begin{equation}
\mathrm{H}(t)\bm{b}^{(1,j)}(t)=\hbar\lambda_{j}(t)\bm{b}^{(1,j)}(t).
\label{eigen-eq}%
\end{equation}
Simple calculations bring us to the eigenstates
\begin{equation}
\bm{b}^{(1,-)}={\frac{\sqrt{2}}{2}}\left(\psi_{1}
\sin\vartheta-\psi_{2}+\psi_{3}\cos\vartheta\right),\label{}\label{eigen-vec-m}
\end{equation}
\begin{equation}
\bm{b}^{(1,0)}=\psi_{1}\cos\vartheta-\psi_{3}\sin\vartheta,\label{eigen-vec-0}
\end{equation}
\begin{equation}
\bm{b}^{(1,+)}={\frac{\sqrt{2}}{2}}\left(\psi_{1}\sin\vartheta+\psi_{2}+\psi_{3}\cos\vartheta\right)
\label{eigen-vec-p}
\end{equation}
and the corresponding eigenvalues of Hamiltonian
\begin{equation}
\hbar\lambda_{1,\pm}=\pm\frac{1}{2}\hbar\Omega,\quad\hbar\lambda_{1,0}=0.
\label{eigen-val}%
\end{equation}
Hereafter, to make notations short, we do not indicate the dependences of
$\vartheta$, $\Omega$, $\bm{b}^{(1,j)}$, $\lambda_{1,j}$ ($j=0,\pm$), and the
vectors of the rotating basis $\psi_{n}$ ($n=1,2,3$) on time (of course, if it
does not cause misunderstanding). Index~1 in the notations $\bm{b}^{(1,j)}$
and $\lambda_{1,j}$ is introduced for the convenience of subsequent
calculations. It means the order of the adiabatic basis [further, we
consider the adiabatic bases of higher (second and third) orders].

The state $\bm{b}^{(1,0)}$ is known from the literature as a \textquotedblleft
dark\textquotedblright\ one, because an atom does not emit light from it
\cite{Fle96,Vit01}.

Provided that the sequence of pulses is \textquotedblleft
counterintuitive\textquotedblright, i.e. when the atom is first
subjected to the action of the Stokes pulse and then, with a certain
delay, of the pumping one which continues to affect the atom for
some time after the Stokes pulse terminates, the angle $\vartheta$
changes, according to formula (\ref{mix-angle}), from zero to
$\pi/2$. Taking the initial conditions (\ref{ini}) into account, we
see that the atom is in the adiabatic state $\bm{b}^{(1,0)}$ at the
beginning of its interaction with the pumping field. If the
parameters of light pulses change slowly enough, the atom remains in
this adiabatic state during the whole period of the simultaneous
interaction with the fields of both pulses. At the moment $t_{f}$,
when the Stokes field is switched off, $\vartheta=\pi/2$. As is seen
from expression (\ref{eigen-vec-0}), $\bm{b}^{(1,0)}$ coincides with
the state $\psi_{3}$ in this case, i.e. the atom transits from state
$\left\vert 1\right\rangle $ into state $\left\vert 3\right\rangle $
due to its interaction with the field. The probability of population
transfer $\left\vert 1\right\rangle \rightarrow\left\vert
3\right\rangle $ is close to 1, provided that the process of
interaction between the atom and the field is close to the adiabatic
one. The stay of an atom or a molecule in the state $\bm{b}^{(1,0)}$
during the whole period of its interaction with the field composes
the physical basis of STIRAP.

Using the basis of adiabatic states, the wave function can be written down in
the form
\begin{equation}
\Psi=\sum\limits_{j=0,\pm}a_{1,j}(t)\bm{b}^{(1,j)}(t). \label{Psi-ad}%
\end{equation}
Here, $a_{1,j}(t)$ is the probability amplitude of finding the atom in the
$j$-th adiabatic state. Substituting function (\ref{Psi-ad}) into the
Schr\"{o}dinger equation (\ref{Sch}), we obtain the Schr\"{o}dinger equation
in the adiabatic basis. In the matrix representation, the state vector looks
like $[a_{1,-}(t),a_{1,0}(t),a_{1,+}(t)]^{T}$, and the Hamiltonian like
\begin{equation}
\mathrm{H}^{(1)}(t)=\frac{\hbar}{2}\left[
\begin{array}
[c]{ccc}%
-\Omega & i\sqrt{2}\dot{\vartheta} & 0\\
-i\sqrt{2}\dot{\vartheta} & 0 & -i\sqrt{2}\dot{\vartheta}\\
0 & i\sqrt{2}\dot{\vartheta} & \Omega\\
&  &
\end{array}
\right]  . \label{Ham-ad}%
\end{equation}
The nonadiabaticity of the atom-field interaction is described by non-diagonal
elements of Hamiltonian (\ref{Ham-ad}). Taking into account that
$\dot{\vartheta}\sim1/\tau$, we see that non-diagonal elements are about $\varepsilon\Omega$ by the order
of magnitude. If they are neglected, the vector
of atomic state in the adiabatic approximation looks as
\begin{equation}
\Psi=\sum\limits_{j=0,\pm}a_{1,j}(t_{i})\bm{b}^{(1,j)}(t)\exp\Biggl(-i\int
\limits_{t_{i}}^{t}\lambda_{1,j}(t^{\prime})dt^{\prime}\Biggr).
\label{Psi-adiab}%
\end{equation}
One can see that, to within the phase, the amplitudes of adiabatic states
remain constant during the whole period of the atom-field interaction. If
the atom was in the \textquotedblleft dark\textquotedblright\ state
$\bm{b}^{(1,0)}$ at the beginning of its interaction with the pumping pulse
(the time moment $t_{i}$), it stays in it after the interaction terminates.
Hence, in the adiabatic approximation, we have the population transfer between
states $\left\vert 1\right\rangle $ and $\left\vert 3\right\rangle $ with the
probability equal to 1.

Now, let us find small nonadiabaticity-induced corrections to the population
transfer from state $\left\vert 1\right\rangle $ to state $\left\vert
3\right\rangle $ in the general case of an arbitrary pulse shape. For this
purpose, it is necessary to take into consideration that the derivative of $\vartheta
$ in Hamiltonian (\ref{Ham-ad}) differs from zero. Let us take
advantage of the formalism used for the description of the quantum-mechanical
system in adiabatic bases of higher orders, which is well-known from the
literature (see \cite{Elk95,Lim91,Fle98}). Let us pass to the basis of eigenstates
of Hamiltonian (\ref{Ham-ad})
\begin{equation}
\bm{b}^{(2,-)}=\frac{\tilde{\Omega}+\Omega}{2\tilde{\Omega}}\bm{b}^{(1,-)}+\frac{i\sqrt{2}\dot\vartheta}{\tilde{\Omega}}\bm{b}^{(1,0)}+\frac{\tilde{\Omega}-\Omega}{2\tilde{\Omega}}\bm{b}^{(1,+)},\label{eigen-vec-ma}
\end{equation}
\begin{equation}
\bm{b}^{(2,0)}=-\frac{\sqrt{2}\dot\vartheta}{\tilde{\Omega}}\bm{b}^{(1,-)}+\frac{i\Omega}{\tilde{\Omega}}\bm{b}^{(1,0)}+\frac{\sqrt{2}\dot\vartheta}{\tilde{\Omega}}\bm{b}^{(1,+)},\label{eigen-vec-0a}
\end{equation}
\begin{equation}
\bm{b}^{(2,+)}=\frac{\Omega-\tilde{\Omega}}{2\tilde{\Omega}}\bm{b}^{(1,-)}+\frac{i\sqrt{2}\dot\vartheta}{\tilde{\Omega}}\bm{b}^{(1,0)}-\frac{\tilde{\Omega}+\Omega}{2\tilde{\Omega}}\bm{b}^{(1,+)}
\label{eigen-vec-pa}
\end{equation}
with corresponding eigenvalues
\begin{equation}
\hbar\lambda_{2,\pm}=\pm\frac{1}{2}\tilde{\Omega},\quad\hbar\lambda_{2,0}=0.
\label{eigen-val-a}%
\end{equation}
In Eqs.~(\ref{eigen-vec-ma})--(\ref{eigen-val-a}), we introduced the notation
\begin{equation}
\tilde{\Omega}=\sqrt{\Omega^{2}+4\dot{\vartheta}^{2}}. \label{eigen-a}%
\end{equation}

In the basis of adiabatic states $\bm{b}^{(2,j)}$, the wave function can be
written down in the form analogous to expression (\ref{Psi-ad}) with the
substitution $1\rightarrow2$
\begin{equation}
\Psi=\sum\limits_{j=0,\pm}a_{2,j}(t)\bm{b}^{(2,j)}(t). \label{Psi-ad-a}%
\end{equation}
Here, $a_{2,j}(t)$ is the probability amplitude of finding the atom in the
state $\bm{b}^{(2,j)}$. In the matrix representation, the state vector in the
basis of states $\bm{b}^{(2,j)}$ looks like $[a_{2,-}(t),a_{2,0}%
(t),a_{2,+}(t)]^{T}$, and the Hamiltonian like
\begin{equation}
\mathrm{H}^{(2)}(t)=\frac{\hbar}{2}\left[
\begin{array}
[c]{ccc}%
-\tilde{\Omega} & i\beta & 0\\
-i\beta & 0 & -i\beta\\
0 & i\beta & \tilde{\Omega}\\
&  &
\end{array}
\right]  , \label{Ham-ad-a}%
\end{equation}
where the notation
\begin{equation}
\beta=\frac{{2\sqrt{2}}}{\tilde{\Omega}^{2}}\left(  \Omega\ddot{\vartheta
}-\dot{\Omega}\dot{\vartheta}\right)  \label{beta}%
\end{equation}
was used. Neglecting the non-diagonal elements in Hamiltonian (\ref{Ham-ad-a}%
) -- or, equivalently, the dependence of $a_{2,j}(t)$ on time, -- we
obtain the wave function of an atom in the form of a superposition
of characteristic states of Hamiltonian (\ref{Ham-ad}):
\begin{equation}
\Psi=\sum\limits_{j=0,\pm}a_{2,j}(t_{i})\bm{b}^{(2,j)}(t)\exp\Biggl(-i\int
\limits_{t_{i}}^{t}\lambda_{2,j}(t^{\prime})dt^{\prime}\Biggr).
\label{Psi-adiab-a}%
\end{equation}
Passing from the basis $\bm{b}^{(2,j)}(t)$ to the basis $\bm{b}^{(1,j)}%
(t)$ and, then, to $\psi_{n}$ with the use of relations (\ref{eigen-vec-m}%
)--(\ref{eigen-vec-p}) and (\ref{eigen-vec-ma})--(\ref{eigen-vec-pa}), we find
the population amplitudes $c_{n}(t)$ for states $\psi_{n}$:
\[
c_{1}(t)=a_{2,-}(t_{i})\left(\frac{\sqrt{2}}{2}\sin\vartheta+\frac{i\sqrt{2}\dot\vartheta\cos\vartheta}{\tilde{\Omega}}\right)e^{i\Phi(t)}-\]
\[-a_{2,+}(t_{i})\left(\frac{\sqrt{2}}{2}\sin\vartheta-\frac{i\sqrt{2}\dot\vartheta\cos\vartheta}{\tilde{\Omega}}\right)e^{-i\Phi(t)}+
\]
\begin{equation}
+a_{2,0}(t_{i})\frac{i\Omega}{\tilde{\Omega}}\cos\vartheta,
\label{c1}
\end{equation}
\[c_{2}(t)=-a_{2,-}(t_{i})\frac{\sqrt{2}\Omega}{2\tilde{\Omega}}e^{i\Phi(t)}+a_{2,0}(t_{i})\frac{2\dot\vartheta}{\tilde{\Omega}}+\]
\begin{equation}
-a_{2,+}(t_{i})\frac{\sqrt{2}\Omega}{2\tilde{\Omega}}\sin\vartheta{}e^{-i\Phi(t)},
\label{c2}
\end{equation}
\[c_{3}(t)=a_{2,-}(t_{i})\left(\frac{\sqrt{2}}{2}\cos\vartheta-\frac{i\sqrt{2}\dot\vartheta\sin\vartheta}{\tilde{\Omega}}\right)e^{i\Phi(t)}-\]
\[-a_{2,+}(t_{i})\left(\frac{\sqrt{2}}{2}\cos\vartheta+\frac{i\sqrt{2}\dot\vartheta\sin\vartheta}{\tilde{\Omega}}\right)e^{-i\Phi(t)}-\]
\begin{equation}
-a_{2,0}(t_{i})\frac{i\Omega\sin\vartheta}{\tilde{\Omega}},
\label{c3}
\end{equation}
where
\begin{equation}
\Phi(t)=\frac{1}{2}\int\limits_{t_{i}}^{t}\tilde{\Omega}(t^{\prime}%
)dt^{\prime}. \label{Phi-2}%
\end{equation}

\begin{figure}
\includegraphics{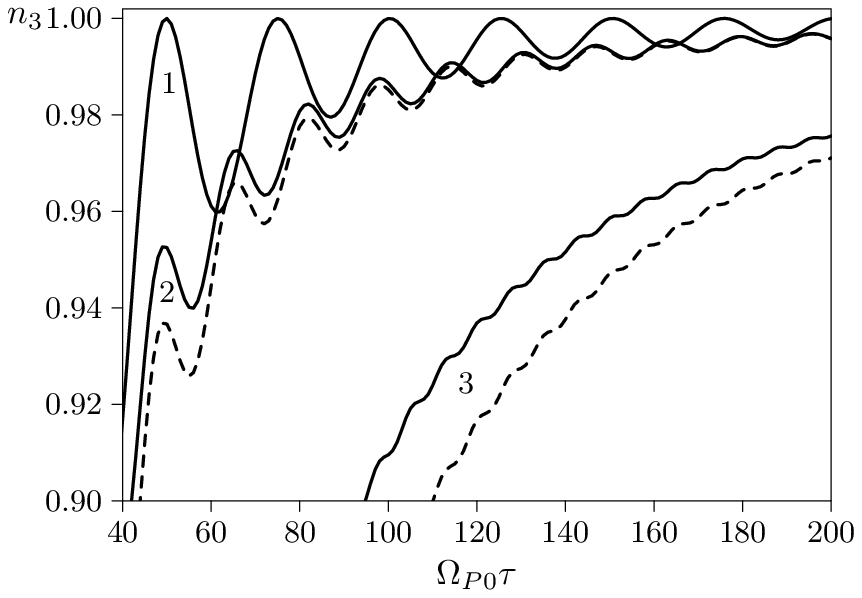}
\vskip-3mm\caption{Dependences of the probability of population
transfer $n_{3}$ from the atomic state $\left\vert 1\right\rangle $
into state $\left\vert 3\right\rangle $ on the Rabi frequency of a
pumping pulse $\Omega_{P0}$ measured in $1/\tau$-units in the field
of light pulses of form (\ref{envelope}), (\ref{cos}) with $n=1$ and
$t_{d}=\tau/2$ for various ratios between $\Omega_{P0}$ and
$\Omega_{S0}$, calculated by formula (\ref{n3}) and by the numerical
integration of the Schr\"{o}dinger equation (\ref{Sch}) with
Hamiltonian (\ref{Ham}). For curve \textit{1} with
$\Omega_{P0}=\Omega_{S0}$, both calculation methods give identical
results. Curves \textit{2} were calculated
for $\Omega_{S0}=2\Omega_{P0}$, curves \textit{3} for $\Omega_{S0}%
=5\Omega_{P0}$. Dashed curves denote the results of numerical
integration of the Schr\"{o}dinger equation   }
\end{figure}

\noindent The coefficients $a_{2,j}(t_{i})$ are determined from
the initial conditions (\ref{ini})
\begin{equation}
a_{2,-}(t_{i})=a_{2,+}(t_{i})=-\frac{i\sqrt{2}\dot\vartheta(t_{i})}{\tilde{\Omega}(t_{i})},\label{apm}
\end{equation}
\begin{equation}
a_{2,0}(t_{i})=-\frac{i\Omega(t_{i})}{\tilde{\Omega}(t_{i})}.\label{a0}
\end{equation}

After the Stokes pulse terminates, the population of state
$\left\vert 3\right\rangle $ does not change any more in time.
From formulas (\ref{c1})--(\ref{a0}), we find that
\[
n_{3}=\left[\Omega(t_{i})\Omega(t_{f})+4\dot\vartheta(t_{i})\dot\vartheta(t_{f})\cos\int\limits_{t_{i}}^{t_{f}}\frac{\tilde{\Omega}(t)}{2}dt)\right]^{2}\times\]
\begin{equation}
\times\tilde{\Omega}(t_{i})^{-2}\tilde{\Omega}(t_{f})^{-2}.
\label{n3}
\end{equation}

According to the result obtained, the probability $n_{3}=|c_{3}|^{2}$ of
population transfer $\left\vert 1\right\rangle \rightarrow\left\vert
3\right\rangle $ is governed by the first derivative of the mixing angle
$\vartheta$ (formula (\ref{mix-angle})) at the beginning of a pumping pulse
(time $t_{i}$)
\begin{equation}
\dot{\vartheta}(t_{i})=\frac{\dot{\Omega}_{P}(t_{i})}{\Omega_{S}(t_{i})}.
\label{mix-ti}%
\end{equation}
and at the end of a Stokes pulse (time $t_{f})$
\begin{equation}
\dot{\vartheta}(t_{f})=-\frac{\dot{\Omega}_{S}(t_{i})}{\Omega_{P}(t_{i})}.
\label{mix-tf}%
\end{equation}
From formulas (\ref{n3})--(\ref{mix-tf}), it follows that $n_{3}$
reaches its maximal value, if
\begin{equation}
\dot{\Omega}_{P}(t_{i})=\pm\dot{\Omega}_{S}(t_{f}). \label{opt}%
\end{equation}
Condition (\ref{opt}) shows that the optimum conditions for population transfer
are obtained, in particular, for symmetric, with respect to the maximum,
pulses with identical shapes and amplitudes. In this case, provided that the
pulse amplitude or the pulse delay is selected properly, so that the cosine in
Eq.~(\ref{n3}) is equal to 1, the population transfer from state
$\left\vert 1\right\rangle $ into state $\left\vert 3\right\rangle $ is
complete in the approximation of adiabatic atomic evolution in the basis of
states $\bm{b}^{(2,j)}(t)$. At the same time, the deviation of $n_{3}$ from
1 with the variation of the cosine argument can reach $16\dot{\vartheta}%
(t_{i})^{2}/\Omega(t_{i})^{2}$, which is of the order of $\varepsilon^{2}$.

Note that, provided that the mixing angle is proportional to time during the
simultaneous action of light pulses on the atom and the frequency $\Omega$
does not depend on time, the non-diagonal elements in Hamiltonian
(\ref{Ham-ad-a}) are equal to zero, and functions (\ref{c1})--(\ref{Phi-2}) are the
exact solutions of the Schr\"{o}dinger equation \cite{Lai96}.

In Fig.~2, the results calculated by formula (\ref{n3}) for pulses with
envelope (\ref{envelope}), (\ref{cos}) with $n=1$ and $t_{d}=\tau/2$ are
depicted for various ratios between $\Omega_{P0}$ and $\Omega_{S0}$. They are
also compared with the results of numerical integration of the Schr\"{o}dinger
equation (\ref{Sch}) with Hamiltonian (\ref{Ham}) and the same parameters of
the atom-field interaction. For curve \textit{1} corresponding to $\Omega
_{P0}=\Omega_{S0}$, both calculation methods give an identical result, because
expression (\ref{n3}) is the exact solution of the Schr\"{o}dinger equation in
this case. As is seen from the figure, the probability of population transfer
tends to 1 with increase of $\Omega_{P0}\tau$, which corresponds to
a reduction of the adiabaticity parameter $\varepsilon$. Simultaneously, the
discrepancy between the results of calculations by formula (\ref{n3}) and
numerical integration of the Schr\"{o}dinger equation decreases.

Equation (\ref{n3}) includes the first derivatives of the mixing angle
$\vartheta$ with respect to time calculated at the beginning of the pumping pulse
and the end of the Stokes one. If the derivatives are equal to zero, then
$n_{3}=1$. In this case, in order to find an expression for $n_{3}$, which
would make allowance for the nonadiabaticity of the atom-field interaction, we
should seek the wave function in the adiabatic basis of higher order than that of
$\bm{b}^{(2,j)}(t)$.

Now, let us analyze how the non-zero second derivatives of
$\vartheta$ at the beginning of the pumping pulse and at the end of
the Stokes one -- either or both -- affect the probability of
population transfer between states $\left\vert 1\right\rangle $ and
$\left\vert 3\right\rangle $ under conditions close to those of the
adiabatic atom-field interaction. The calculation routine is the
same, as was used at the derivation of formula (\ref{n3}) for the
population of state $\left\vert 3\right\rangle $. Being interested
only in the case where the proximity of the population transfer to
the adiabatic one is governed by the second derivative of the
pumping field at the beginning of the pumping pulse and the second
derivative of the Stokes field at the end of the Stokes pulse, we
assume that
\begin{equation}
\dot{\vartheta}(t_{i})=\dot{\vartheta}(t_{f})=0.\label{ini-2}%
\end{equation}
We obtain the eigenvalues $\hbar\lambda_{3,j}$ ($j=\pm,0$) and eigenstates
$\bm{b}^{(3,j)}(t)$ of Hamiltonian (\ref{Ham-ad-a}) and take them as a new
basis. In the Hamiltonian $H^{(3)}(t)$ which describes the atom-field
interaction in this basis, we neglect non-diagonal elements. In this basis,
the wave function looks like (\ref{Psi-adiab-a}) with index 2 being
substituted by 3. With regard for the relations between basis vectors
$\bm{b}^{(3,j)}(t)$, $\bm{b}^{(2,j)}(t)$, $\bm{b}^{(1,j)}(t)$, and $\psi_{n}$
($n=1,2,3$), as well as the initial conditions (\ref{ini}), we find the population
of state $\left\vert 3\right\rangle $:
\begin{equation}
n_{3}=\frac{\left[  \Omega(t_{i})^{2}\Omega(t_{f})^{2}+16\ddot{\vartheta
}(t_{i})\ddot{\vartheta}(t_{f})\cos\tilde{\Phi}\right]  ^{2}}{\left(
\Omega(t_{i})^{4}+16\ddot{\vartheta}(t_{i})^{2}\right)  \left(  \Omega
(t_{f})^{4}+16\ddot{\vartheta}(t_{f})^{2}\right)  },\label{n3-2}%
\end{equation}
where
\begin{equation}
\tilde{\Phi}=\frac{1}{2}\int\limits_{t_{i}}^{t_{f}}\sqrt{\tilde{\Omega}%
^{2}+16\left(  \dot{\vartheta}\dot{\Omega}-\Omega\ddot{\vartheta}\right)
^{2}\tilde{\Omega}^{-4}}dt.\label{Phi-a}%
\end{equation}

The second derivatives of the mixing angle at the beginning of the pumping pulse and at the
end of the Stokes one are coupled with the second derivatives of the corresponding
Rabi frequencies,
\begin{equation}
\ddot{\vartheta}(t_{i})=\frac{\ddot{\Omega}_{P}(t_{i})}{\Omega_{S}(t_{i}%
)},\qquad\ddot{\vartheta}(t_{f})=-\frac{\ddot{\Omega}_{S}(t_{f})}{\Omega
_{P}(t_{f})}. \label{dd_theta}%
\end{equation}
From expression (\ref{n3-2}), it follows that $n_{3}$ reaches the
maximal value, when
\begin{equation}
\Omega_{P}(t_{i})^{2}\ddot{\vartheta}(t_{f})=\pm\Omega_{S}(t_{f})^{2}%
\ddot{\vartheta}(t_{i}). \label{opt-2}%
\end{equation}

\noindent As is seen from formula (\ref{opt-2}), the maximal
population transfer between states $\left\vert 1\right\rangle $ and
$\left\vert 3\right\rangle $ is achieved, in particular, in the
field of pulses, for which the values of $|\ddot{\vartheta}|$ are
identical at the beginning of the pumping pulse and at the end of the Stokes
one, $|\ddot{\vartheta}(t_{i})|=|\ddot{\vartheta}(t_{f})|$, and
the Rabi frequencies are also identical at these time moments,
$\Omega (t_{i})=\Omega(t_{f})$. In this case, provided that the
intensities of pulses are selected properly, so that $\tilde{\Phi}=2\pi
n$, where $n$ is an integer number, the probability of population
transfer is close to 1. The deviation of the probability from 1
with the variation of $\tilde{\Phi}$ is about $\varepsilon^{4}$ by
an order of magnitude.

\begin{figure}
\includegraphics{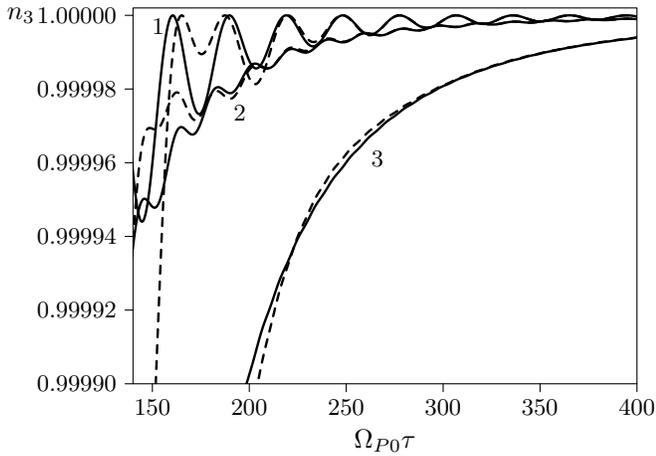}
\vskip-3mm\caption{Dependences of the probability of population
transfer $n_{3}$ from atomic state $\left\vert 1\right\rangle $ into
state $\left\vert 3\right\rangle $ on the Rabi frequency of a
pumping pulse $\Omega_{P}$ measured in $1/\tau$-units in the field
of light pulses of form (\ref{envelope}), (\ref{cos}) with $n=2$ and
$t_{d}=\tau/2$ for various ratios between $\Omega_{P0}$ and
$\Omega_{S0}$, calculated by formula (\ref{n3-2}) and by the
numerical integration of the Schr\"{o}dinger equation (\ref{Sch})
with Hamiltonian (\ref{Ham}). For curves \textit{1},
$\Omega_{P0}=\Omega_{S0}$; curves \textit{2} were calculated for
$\Omega_{S0}=2\Omega_{P0}$, and curves \textit{3} for
$\Omega_{S0}=5\Omega_{P0}$. Dashed curves denote the results of
numerical integration of the Schr\"{o}dinger equation  }
\end{figure}

In Fig.~3, the results of calculations by formula (\ref{n3-2}) for pulses of
form (\ref{envelope}),(\ref{cos}) with $n=2$ and $t_{d}=\tau/2$ for various
ratios between $\Omega_{P0}$ and $\Omega_{S0}$ are depicted and compared with
the results obtained by numerical integration of the Schr\"{o}dinger equation
(\ref{Sch}) with Hamiltonian (\ref{Ham}) and the same parameters of
the atom-field interaction. As is seen, the maximal values of $n_{3}$ are reached
for pulses with $|\ddot{\vartheta}(t_{i})|=|\ddot{\vartheta}(t_{f})|$ and
$\Omega(t_{i})=\Omega(t_{f})$, in accordance with the analysis of expression
(\ref{n3-2}) given above for the probability of population transfer. The
oscillations of the population expectedly decrease with increase of the pulse area,
i.e. as the interaction with the field approaches the adiabatic one.

The obtained results [formulas (\ref{n3}) and (\ref{n3-2})] describe the
dependence of the population transfer on the light pulse parameters in the case
where the interaction between the atom and the field is close to the adiabatic one. As
is seen from Figs.~2 and 3, they correctly describe the oscillating dependence
of the population transfer probability in this range of parameters. In Fig.~3 which
corresponds to nonzero second derivatives of the mixing angle at the moments
$t_{i}$ and $t_{f}$, the plotted curves are located much closer to the
value $n_{3}=1$, than the corresponding curves in Fig.~2, where already the
first derivatives of the mixing angle are different from zero. It was to be
expected, because the amplitudes of oscillations are of the order of
$\varepsilon^{4}$ and $\varepsilon^{2}$ for the dependences plotted in Figs.~3
and 2, respectively.

If both the first and second derivatives of the mixing angle equal zero at
the time moments $t_{i}$ and $t_{f}$, expressions (\ref{n3}) and (\ref{n3-2})
do not contain any more nonadiabaticity corrections to the probability of
population transfer, and they give $n_{3}=1$. To find these corrections, it is
necessary to pass to the adiabatic basis of higher order. Should the
first derivative different from zero at the moments $t_{i}$ and $t_{f}$ be a
derivative of the $n$-th order, the order of magnitude of the nonadiabaticity
correction would be $\varepsilon^{2n}$.

\section{Stimulated Raman Adiabatic Passage in the Field of Long Light Pulses}

Now, consider the population transfer between atomic states $\left\vert
1\right\rangle $ and $\left\vert 3\right\rangle $ in the course of STIRAP
in the case of long light pulses, when the time of the atom-field
interaction considerably exceeds the time of the spontaneous emission
from the excited state, $\gamma\tau\gg1$. Consider the \textquotedblleft
bright\textquotedblright, $\psi_{b}$, excited, $\psi_{e}$, and
\textquotedblleft dark\textquotedblright, $\psi_{d},$ states defined by
formulas \cite{Fle96}
\begin{equation}
\psi_b=\sin\vartheta(t)\psi_1+\cos\vartheta(t)\psi_3,\label{b}
\end{equation}
\begin{equation}
\psi_e=\psi_2,\label{e}
\end{equation}
\begin{equation}
\psi_d=\cos\vartheta(t)\psi_1-\sin\vartheta(t)\psi_3,\label{d}
\end{equation}
where $\psi_{j}$ ($j=1,2,3$) are the basis wave functions of the rotating basis,
in which Hamiltonian (\ref{Ham}) is written down. The functions $\psi_{j}$
differ from the functions $|j\rangle$ only by time-dependent phases. Only one
of those states, $\psi_{d}$, which coincides with (\ref{eigen-vec-0}), is the
eigenstate of Hamiltonian (\ref{Ham}). At the beginning of the atom-field
interaction, $\vartheta=0$, and the atom is in the state $\psi_{d}$. Should it
stay in this state during the whole period of the interaction with the field,
then, at the moment, when the Stokes pulse terminates and $\vartheta$ grows up
to $\vartheta=\pi/2$, the population would be completely transferred from the
state $\psi_{1}$ into the state $\psi_{3}$.

The vector of state constructed of the probability amplitudes $C_{b}$,
$C_{e}$, and $C_{d}$ to find the atom in the states $\psi_{b}$, $\psi_{e}$,
and $\psi_{d}$, respectively, looks like $[C_{b},C_{e},C_{d}]^{T}$, and the
Hamiltonian in this basis is
\begin{equation}
\mathrm{H}^{(\rm bed)}(t)=\frac{\hbar}{2}\left[
\begin{array}
[c]{ccc}%
0 & \Omega & i\dot{\vartheta}\\
\Omega & -i\gamma & 0\\
-i\dot{\vartheta} & 0 & 0\\
&  &
\end{array}
\right]  . \label{Ham-bed}%
\end{equation}
The probability of population transfer from the state $\psi_{1}$
into the state $\psi_{3}$ -- or, equivalently, the population
$n_{3}$ of the state $\psi_{3}$, which we are interested in -- is
equal to
\begin{equation}
n_{3}=\left\vert C_{3}(t_{f})\right\vert ^{2}. \label{n3-gt}%
\end{equation}

If STIRAP occurs in the field of long pulses, the assumption is
usually made that the characteristic time of a probability amplitude
variation is of the same order of magnitude as the duration of light
pulses \cite{Fle96,Rom97,Yat02}. This assumption allows the
Schr\"{o}dinger equation to be solved by the iteration method,
supposing that the derivatives of amplitudes are small in comparison
with $\Omega$. In essence, the transient processes that start at the
beginning $t_{i}$ of the interaction between the atom and the
pumping pulse are neglected. This approach always gives rise to the
correct first term in the expansion of the probability in a series in
the adiabaticity parameter~$\varepsilon$.

In order to take the correction to the population transfer probability
associated with the transient processes arising at the switching-on of a pumping pulse
into account, we solve the Schr\"{o}dinger equation in two
stages. First, we suppose that the left-hand side of the Schr\"{o}dinger
equation with Hamiltonian (\ref{Ham-bed}) represented in the basis of states
$\psi_{b}$, $\psi_{e}$, and $\psi_{d}$ has the same order of magnitude as the
term proportional to $\Omega$ on the right-hand side, and solve the equation
in the time interval $[t_{i},t_{1}]$, where $\gamma^{-1}\ll t_{1}-t_{i}\ll
\tau$. Such a $t_{1}$-value can always be found, bearing in mind the condition
of long interaction between the atom and the field, $\gamma\tau\gg1$. When
solving the Schr\"{o}dinger equation in this time interval, we neglect the
time dependence $\Omega(t)$ and adopt that $\Omega(t)=\Omega(t_{i})$. As a
result, we take damped oscillations of the amplitudes at the
beginning of the atom-field interaction into account. Further, we solve the Schr\"{o}dinger
equation in the interval $[t_{1},t_{f}],$ by supposing now, as was done in works
\cite{Fle96,Rom97,Yat02}, that the characteristic time of amplitude derivative
variations has an order of magnitude of $\tau$.

Let us pass to the variables
\begin{equation}
\eta_{d}=\ln{C_{d}},\qquad\eta_{b}=C_{b}/C_{d},\qquad\eta_{b}=C_{b}/C_{d}.
\label{eta}%
\end{equation}
From the Schr\"{o}dinger equation (\ref{Sch}) with Hamiltonian
(\ref{Ham-bed}), we find
\begin{equation}
\dot\eta_{b}=-\frac{i}{2}\Omega\eta_{e}-\eta_{b}\dot\eta_{d}+\dot\vartheta,\label{eb}
\end{equation}
\begin{equation}
\dot\eta_{e}=-\frac{i}{2}\Omega\eta_{b}-\frac{1}{2}\gamma\eta_{e}-\eta_{e}\dot\eta_{d},\label{ee}
\end{equation}
\begin{equation}
\dot\eta_{d}=-\eta_{b}\dot\vartheta.\label{ed}
\end{equation}
The population of state $\left\vert 3\right\rangle $ after the Stokes pulse
terminates coincides with the population in the state $\psi_{d}$, being equal
to
\begin{equation}
n_{3}=\exp\left[  {2\eta_{d}(t_{f})}\right]  =\exp{\int\limits_{t_{i}}^{t_{f}%
}2\dot{\eta}_{d}(t)dt}. \label{n3-g}%
\end{equation}

Consider Eqs.~(\ref{eb})--(\ref{ed}) in the time interval $[t_{i},t_{1}]$,
where $\vartheta$ is small. To mark this smallness, let us formally introduce
the parameter $\epsilon\ll1$ at $\vartheta$ in those equations (at
the end of calculations, we put $\epsilon=1$) and seek $\eta_{b}$,
$\eta_{e}$, and $\dot{\eta}_{d}$ in the form
\begin{equation}
\eta_{b}=\sum\limits_{n=0}^{\infty}H_{b,n}\epsilon^{n},\label{Eta-b}
\end{equation}
\begin{equation}
\eta_{e}=\sum\limits_{n=0}^{\infty}H_{e,n}\epsilon^{n},\label{Eta-e}
\end{equation}
\begin{equation}
\dot\eta_{d}=\sum\limits_{n=0}^{\infty}H_{d,n}\epsilon^{n}.\label{Eta-d}
\end{equation}
First, let us consider the case $\dot{\vartheta}(t_{i})\not =0$, i.e. when the Rabi
frequency of a pumping pulse at its beginning linearly depends on time (see
Eq.~(\ref{mix-ti})). Substituting Eqs.~(\ref{Eta-b})--(\ref{Eta-d}) in
Eqs.~(\ref{eb})--(\ref{ed}) with regard for the initial conditions
\begin{equation}
\eta_{b}=\eta_{e}=\eta_{b}=0, \label{ini-eta}%
\end{equation}
which follow from Eq.~(\ref{ini}), we find, after simple calculations, that
\begin{equation}
H_{d,0}    =H_{d,1}=0,\label{Hd}\end{equation}
\[
H_{d,2}    =-\frac{2\gamma\alpha^{2}}{\Omega_{i}^{2}}+\frac{2\alpha^{2}%
\gamma}{\Omega_{i}^{2}}e^{-\frac{1}{4}\gamma{}t^{\prime}}\cos\!\left(
\!\frac{t^{\prime}}{2}\sqrt{\Omega_{i}^{2}-{\textstyle\frac{1}{4}}\gamma^{2}%
}\right)  +\]
 \begin{equation}+\frac{\alpha^{2}\left(  \gamma^{2}-2\Omega_{i}%
^{2}\right)  e^{-\frac{1}{4}\gamma{}t^{\prime}}}{\Omega_{i}^{2}\sqrt
{\Omega_{i}^{2}-{\textstyle\frac{1}{4}}\gamma^{2}}}\sin\!\left(
\!\frac{t^{\prime}}{2}\sqrt{\Omega_{i}^{2}-{\textstyle\frac{1}{4}}\gamma^{2}%
}\right)  , \label{Hd2}%
\end{equation}
which is necessary for further calculations of the population in the
state $\psi_{d}$. Here, the notations
\begin{equation}
\Omega_{i}=\Omega(t_{i}),\qquad\alpha=\dot{\vartheta}(t_{i}),\qquad{}%
t^{\prime}=t-t_{i}. \label{not-H}%
\end{equation}
are used.

Now, consider the time interval $[t_{1},t_{f}]$, where oscillations of
the population in the states $\psi_{b}$, $\psi_{e}$, and $\psi_{d}$ practically
disappear (since $\gamma t_{1}\gg1$), and the amplitudes of those states
slowly change in time and approximately follow the variations of light pulse
Rabi frequencies. Similarly to what was done when solving Eqs.~(\ref{eb}%
)--(\ref{ed}) in the time interval $[t_{i},t_{1}]$, we formally
introduce a small parameter $\epsilon$ into them to mark the
magnitude of coefficients. Since $\max({\Omega}\tau)\gg1$ and
$\gamma\tau\gg1$, let us write down Eqs.~(\ref{eb})--(\ref{ed}) in
the form
\begin{equation}
\dot\eta_{b}=-\frac{i}{2}\Omega\eta_{e}\epsilon^{-1}-\eta_{b}\dot\eta_{d}+\dot\vartheta,\label{ebt}
\end{equation}
\begin{equation}
\dot\eta_{e}=-\frac{i}{2}\Omega\eta_{b}\epsilon^{-1}-\frac{1}{2}\gamma\eta_{e}\epsilon^{-1}-\eta_{e}\dot\eta_{d},\label{eet}
\end{equation}
\begin{equation}
\dot\eta_{d}=-\eta_{b}\dot\vartheta.\label{edt}
\end{equation}
The absence of the factor $\epsilon^{-1}$ at $\dot{\eta}_{b}$, $\dot{\eta}%
_{e}$, and $\dot{\eta}_{d}$ means that the characteristic times of their
variations are of the order of the light pulse length $\tau$.

\begin{figure}
\includegraphics{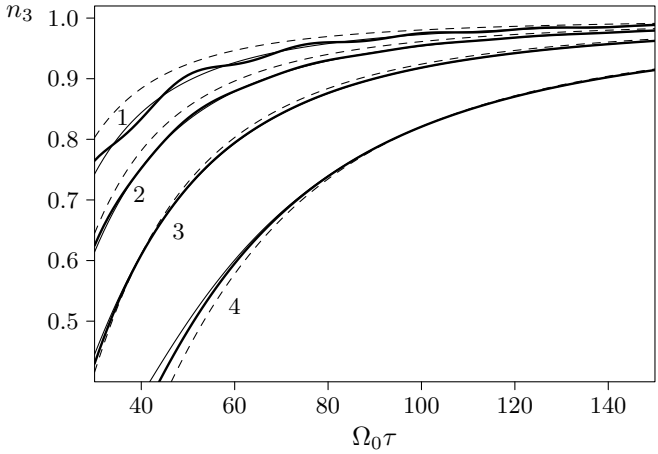}
\vskip-3mm\caption{Dependences of the probability of population
transfer $n_{3}$ from the atomic state $\left\vert 1\right\rangle $
into state $\left\vert 3\right\rangle $ on $\Omega_{0}\tau$ in the
field of light pulses of form (\ref{envelope}), (\ref{cos}) with
$n=1$ and $t_{d}=\tau/2$ calculated for various $\gamma\tau=10$
(\textit{1}), 20 (\textit{2}), 40 (\textit{3}), and 100 (\textit{4})
by formula (\ref{an-cos}) and by the numerical integration of the
Schr\"{o}dinger equation (\ref{Sch}) with Hamiltonian (\ref{Ham}).
Thick solid curves are the results of numerical integration of the
Schr\"{o}dinger equation, thin solid curves are the results of
calculations by formula (\ref{an-cos}), dashed curves are the
results of calculations by the same formula without regard for the
first term in the exponent which is responsible for a
nonadiabaticity that is inserted by the jump of the first derivative
of $\vartheta$ at the beginning of a pumping pulse  }
\end{figure}

The solution of Eqs.~(\ref{ebt})--(\ref{edt}) is sought in form (\ref{Eta-b}%
)--(\ref{Eta-d}). After simple calculations, we obtain
\begin{equation}
H_{d,0}=0,\qquad{}H_{d,1}=-\frac{2\gamma\dot\vartheta^{2}}{\Omega^{2}},
\label{Hd1g}
\end{equation}
\begin{equation}
H_{d,2}=\frac{4\dot\vartheta^{2}\dot\Omega\left(\Omega^{2}-2\gamma^{2}\right)}{\Omega^{5}}+
\frac{4\dot\vartheta\ddot\vartheta\left(\gamma^{2}-\Omega^{2}\right)}{\Omega^{4}}.
\label{Hd2g}
\end{equation}
The expression for $\dot{\eta}_{d}$ obtained from Eqs.~(\ref{Eta-d})
and (\ref{Hd})--(\ref{Hd2}) for $t_{i}\leq t\leq t_{1}$ has to
transform at the time moment $t=t_{1}$ into the corresponding
expression obtained for $t_{1}\leq t\leq t_{f}$. Really, the
quantity $H_{d,1}(t_{1})$ from Eq.~(\ref{Hd1g}) is equal to
$H_{d,2}(t_{1})$ from Eq.~(\ref{Hd2}), because the oscillating terms
in the latter, due to the exponential damping, are practically
zeroed at this moment. No misunderstanding should invoke the
comparison made between terms of different $\epsilon$-orders,
because this parameter is different at different time intervals: in
the former case, it marks terms with the $\dot{\vartheta}$ order of
magnitude; in the latter, the terms of the order of the adiabaticity
parameter $\varepsilon$.

The terms that correspond to $H_{d,2}$ from Eq.~(\ref{Hd2g}) and
have higher $\epsilon$-orders than those presented in
Eqs.~(\ref{Hd})--~(\ref{Hd2}) have to appear in the interval
$[t_{i},t_{1}]$. The indicated order is the maximal one in the
expansion of $\dot{\eta}_{d}$ that should be taken into account for
a linear time dependence of the Rabi frequency of the pumping field,
because the exceeding of the calculation precision occurs otherwise.

Now, let us calculate the probability of population transfer from state
$\left\vert 1\right\rangle $ into state $\left\vert 3\right\rangle $. From
Eqs.~(\ref{n3-g}), (\ref{Hd2}), (\ref{Hd1g}), and (\ref{Hd2g}), we find
\[
n_{3}    =\exp\biggl[\frac{8\alpha^{2}}{\Omega_{i}^{4}}\left( \gamma
^{2}-\Omega_{i}^{2}\right) -4\gamma\int\limits_{t_{i}}^{t_{f}}\frac
{\dot{\vartheta}^{2}}{\Omega^{2}}dt+
\]
\begin{equation}%
+8\int\limits_{t_{i}}^{t_{f}}\!\biggl(\!\frac
{\dot{\vartheta}^{2}\dot{\Omega}\left( \Omega^{2}-2\gamma^{2}\right)
}{\Omega^{5}}+\frac{\dot{\vartheta}\ddot{\vartheta}\left(  \gamma^{2}%
-\Omega^{2}\right)  }{\Omega^{4}}\biggr)dt\biggr].
\label{n3-i}%
\end{equation}

\noindent Here, the first term in the exponent emerges
owing to damped oscillations of the population with a frequency of the
order of $\Omega_{i}$ which arise at the beginning of a pumping
pulse. The other terms obtained in works
\cite{Fle96,Rom97} are associated with the quasistationary
evolution of the populations of atomic states with a
characteristic time of the order of the light pulse duration.

Let us illustrate the obtained result in the case where the integrals in
Eq.~(\ref{n3-i}) can be calculated analytically. Consider light pulses of form
(\ref{envelope}), (\ref{cos}) with $n=1$ and at $\Omega_{P0}=\Omega
_{S0}=\Omega_{0}$ and $t_{d}=\tau/2$. For such pulses,
\begin{equation}
\Omega=\Omega_{0},\qquad\vartheta=\frac{\pi}{\tau}t, \label{analit}%
\end{equation}
and the second integral in Eq.~(\ref{n3-i}) vanishes, because $\dot{\Omega}=0$
and $\ddot{\vartheta}=0$. Simple calculations bring about
\begin{equation}
n_{3}=\exp\biggl[\frac{8\pi^{2}}{\Omega_{0}^{4}\tau^{2}}\left(  \gamma
^{2}-\Omega_{0}^{2}\right)  -\frac{2\gamma\pi^{2}}{\Omega_{0}^{2}\tau}\biggr].
\label{an-cos}%
\end{equation}
As is seen, the relative contribution of expression (\ref{an-cos}) to the
exponent, which is associated with transient processes at the beginning
of a pumping pulse, is of the order of $4/(\gamma\tau)$. For example, at
$\gamma\tau=40$, in the case of the atom-field interaction close to the adiabatic
one, i.e. $\Omega_{0}\tau\gg1$, the corresponding correction to the quantity
$1-n_{3}$ is about 10\%.

In Fig.~4, the results of numerical calculations of the probability
of population transfer from state $\left\vert 1\right\rangle $ into
state $\left\vert 3\right\rangle $ obtained by the numerical
integration of the Schr\"{o}dinger equation are shown, as well as
the results of calculations by formula (\ref{an-cos}), where
allowance is made or not for the first term in the exponent which is
responsible for a nonadiabaticity inserted by the jump in the first
derivative of the mixing angle $\vartheta$ at the beginning of a
pumping pulse. The figure demonstrates that taking the
nonadiabaticity associated with the jump of $\dot{\vartheta}$ at the
time moment $t_{i}$ into consideration substantially improves the
accuracy of $n_{3}$ calculations. The dependence of the population
transfer probability on the light pulse area obtained in such a way
(in Fig.~4, the light pulse area is parametrized by the product
$\Omega_{0}\tau$) practically coincides with the result of numerical
calculations of this quantity from the Schr\"{o}dinger equation at
$\gamma>20/\tau$. The result of calculation by formula
(\ref{an-cos}) at $\gamma\tau=10$ reproduces -- though on the
average -- the corresponding result obtained by the numerical
integration of the Schr\"{o}dinger equation. At the same time, they
appreciably differ from each other, because the condition that there
must be such $t_{1}$, which satisfies both inequalities --
$\gamma{}(t_{1}-t_{i})\gg1$ and $(t_{1}-t_{i})\ll (t_{f}-t_{i})$ --
simultaneously, is violated. Really, in the case of the light pulses
under consideration, $(t_{f}-t_{i})=\tau/2$, so that $\gamma
(t_{1}-t_{i})$ cannot exceed 5. As a result, as is seen from
Eq.~(\ref{Hd2}), the exponent that is responsible for the damping of
population amplitude oscillations in the dark state amounts to only
$e^{-1.25}$ at the end of the simultaneous interaction of the atom
with the fields of Stokes and pumping pulses, which contradicts the
assumption about the oscillation termination within a short, in
comparison with $\tau/2$, time of the atom-field interaction, which
is necessary for expression (\ref{an-cos}) to be valid.\looseness=-1

In the case where the first derivative of the mixing angle $\vartheta$ at the
beginning $t_{i}$ of a pumping pulse is equal to zero, whereas the
second derivative is nonzero, it is also possible, within the calculation
scheme described above, to obtain an expression for the population transfer
probability similar to formula (\ref{n3-i}). The correction in the exponent,
which emerges due to damped population oscillations arising at the beginning
of a pumping pulse, is of the order of $\varepsilon^{4}$ in this case [in
expression (\ref{n3-i}), it is of the order of $\varepsilon^{2}$], which, in
general, is much less than the values of integrals included into
Eq.~(\ref{n3-i}). Hence, it is eligible to neglect the transient processes
arising at the beginning of a pumping pulse, provided that the time dependence
of the mixing angle is described by a power law $\vartheta\sim t^{n}$ with
$n\geq2$. In this case, the probability of population transfer can be found in
the quasistationary approximation, at least with an accuracy of not worse
than $\varepsilon^{3}$, supposing that the characteristic variation times of
atomic state populations are of the order of the light pulse duration
\cite{Fle96,Rom97}.

\section{Conclusions}

We have analyzed the influence of the extra nonadiabaticity
associated with the non-analytical behavior of the field strengths
of light pulses at the beginning and the end of their action upon
the atom in the course of stimulated Raman adiabatic passage on the
probability of population transfer. The cases where the light pulses
are much shorter than the lifetime of an atom in the intermediate
state and when the time of the atom-field interaction considerably
exceeds the time of the spontaneous emission in this state, have
been considered. In both cases, the additional nonadiabaticity is
maximal, when the field strength grows linearly (or the intensity
quadratically) with time at the beginning of a pumping pulse. For
short light pulses, the optimum conditions for population transfer
are reached, if the time-derivative of the Rabi frequency of a
pumping pulse at the time moment of switching-on is equal to that of
a Stokes pulse at the time moment of switching-off. In the case of
long light pulses, the correction, which is related to the transient
processes occurring at the beginning of a pumping pulse, to the
theory developed in works \cite{Fle96,Rom97} can appreciably change
the result only if the first derivative of the pumping field
strength differs from zero.

\vskip3mm The work was executed in the framework of the themes
Nos.~V136 and VTs139 of the NAS of Ukraine.


\begin{thebibliography}{99}                                                                                               %


\bibitem {Sho90}B.W.~Shore, \emph{The Theory of Coherent Atomic Excitation}
(Wiley, New York, 1990).

\bibitem {Sho08}B.W.~Shore, Acta Phys. Slovaka \textbf{58}, 243 (2008).

\bibitem {Ore84}J.~Oreg and F.T.~Hioe, J.H.~Eberly, Phys. Rev.~A \textbf{29},
690 (1984).

\bibitem {Gau88}U.~Gaubatz, P.~Rudecki, M.~Becker, S.~Schiemann, M.~Kulz, and
K.~Bergmann, Chem. Phys. Lett. \textbf{149}, 463 (1988).

\bibitem {Ber98}K.~Bergmann, H.~Theur, and B.W.~Shore, Rev. Mod. Phys.
\textbf{70}, 1003 (1998).

\bibitem {Vit01}N.V.~Vitanov, T.~Halfmann, B.W.~Shore, and K.~Bergmann, Annu.
Rev. Phys. Chem. \textbf{52}, 763 (2001).

\bibitem {Got06}H.~Goto, K.~Ichimura, Phys. Rev.~A \textbf{74}, 053410 (2006).

\bibitem {Alz76}G.~Alzetta, A.~Gozzini, L.~Moi, and G.~Orriols, Nuovo Cimento
B \textbf{36}, 5 (1976).

\bibitem {Ari76}E.~Arimondo and G.~Orriols, Lett. Nuovo Cimento \textbf{17},
333 (1976)

\bibitem {Gra78}H.R.~Gray, R.W.~Whitley, and C.R.~Stroud jr., Opt. Lett.
\textbf{3}, 218 (1978).

\bibitem {Mes79}A.~Messiah, \emph{Quantum Mechanics. Vol.~II }(North-Holland,
Amsterdam, 1962).

\bibitem {Car90}C.E.~Carroll and F.T.~Hioe, Phys. Rev. A \textbf{42}, 1522 (1990).

\bibitem {Lai96}T.A.~Laine and S.~Stenholm, Phys. Rev.~A \textbf{53}, 2501 (1996).

\bibitem {Elk95}M.~Elk, Phys. Rev.~A \textbf{52}, 4017 (1995).

\bibitem {Dyk61}A.M.~Dykhne, Sov. Phys. JETP \textbf{14}, 941 (1962).

\bibitem {Dav76}J.P.~Davis and P.~Pechukas, J.~Chem. Phys. \textbf{64}, 3129 (1976).

\bibitem {Gar62}L.M.~Garrido and F.J.~Sancho, Physica (Amsterdam)
\textbf{28}, 553 (1962).

\bibitem {San66}F.J.~Sancho, Proc. Phys. Soc. \textbf{89}, 1 (1966).

\bibitem {Yat04}L.P.~Yatsenko, S.~Gu\'{e}rin, and H.R.~Jauslin, Phys. Rev.~A
\textbf{70}, 043402 (2004).

\bibitem {Fle96}M.~Fleischhauer and A.S.~Manka, Phys.~Rev.~A \textbf{54}, 794 (1996).

\bibitem {Rom97}V.I.~Romanenko and L.P.~Yatsenko, Opt. Commun. \textbf{140},
231 (1997).

\bibitem {Yat02}L.P.~Yatsenko, V.I.~Romanenko, B.W.~Shore, and K.~Bergmann,
Phys. Rev.~A \textbf{65}, 043409 (2002).

\bibitem {Vit97}N.V.~Vitanov and S.~Stenholm, Phys. Rev.~A \textbf{56}, 1463 (1997).

\bibitem {Lim91}R.~Lim and M.V.~Berry, J. Phys.~A \textbf{24}, 3255 (1991).

\bibitem {Fle98}M.~Fleischhauer, R.~Unanyan, B.W.~Shore, and K.~Bergmann,
Phys. Rev.~A \textbf{59}, 3751 (1998).\vskip3mm

\end{thebibliography}
\end{document}